\documentclass[aps,prl,reprint,amsmath,amssymb,superscriptaddress,longbibliography]{revtex4-1}
\usepackage[utf8]{inputenc}
\usepackage{graphicx}
\usepackage{dcolumn}
\usepackage{bm}
\usepackage{amsmath}
\usepackage{color}
\usepackage{xspace}
\usepackage{ulem}
\usepackage{xfrac}
\usepackage{color,soul}
\usepackage{graphicx}
\usepackage[colorlinks=true,urlcolor=blue,citecolor=blue,linkcolor=blue,bookmarks=false,pdfstartview={FitH}]{hyperref}

\usepackage[bordercolor=white,backgroundcolor=gray!30,linecolor=black,colorinlistoftodos]{todonotes}

\def\BCIO{Ba$_5$CuIr$_3$O$_{12}$\,}
\def\cm-1{cm$^{-1}$\,}

\begin{document}

\title{Random singlet state in Ba$_5$CuIr$_3$O$_{12}$ single crystals}

\author{Pavel A. Volkov}
\email{pv184@physics.rutgers.edu}
\affiliation{Department of Physics and Astronomy, Rutgers University, Piscataway, NJ 08854, USA}
\affiliation{Center for Materials Theory, Rutgers University, Piscataway, New Jersey, 08854, USA}
\author{Choong-Jae Won}
\affiliation{Max Planck POSTECH/Korea Research Initiative, Pohang University of Science and Technology, Pohang 37673, Korea}
\affiliation{Laboratory of Pohang Emergent Materials, Pohang Accelerator Laboratory, Pohang 37673, Korea}
\author{D. I. Gorbunov}
\affiliation{Hochfeld-Magnetlabor Dresden (HLD-EMFL), Helmholtz-Zentrum Dresden-Rossendorf, 01328 Dresden, Germany}
\author{Jaewook Kim}
\affiliation{Department of Physics and Astronomy, Rutgers University, Piscataway, NJ 08854, USA}
\affiliation{Rutgers Center for Emergent Materials, Rutgers University, Piscataway, NJ 08854, USA}
\author{Mai Ye}
\email{mye@physics.rutgers.edu}
\affiliation{Department of Physics and Astronomy, Rutgers University, Piscataway, NJ 08854, USA}
\author{Heung-Sik Kim}
\affiliation{Department of Physics and Astronomy, Rutgers University, Piscataway, NJ 08854, USA}
\affiliation{Department of Physics, Kangwon National University, Chuncheon 24341, Korea}
\author{J. H. Pixley}
\affiliation{Department of Physics and Astronomy, Rutgers University, Piscataway, NJ 08854, USA}
\affiliation{Center for Materials Theory, Rutgers University, Piscataway, New Jersey, 08854, USA}
\author{Sang-Wook Cheong}
\affiliation{Department of Physics and Astronomy, Rutgers University, Piscataway, NJ 08854, USA}
\affiliation{Max Planck POSTECH/Korea Research Initiative, Pohang University of Science and Technology, Pohang 37673, Korea}
\affiliation{Laboratory of Pohang Emergent Materials, Pohang Accelerator Laboratory, Pohang 37673, Korea}
\affiliation{Rutgers Center for Emergent Materials, Rutgers University, Piscataway, NJ 08854, USA}
\author{G. Blumberg}
\email{girsh@physics.rutgers.edu}
\affiliation{Department of Physics and Astronomy, Rutgers University, Piscataway, NJ 08854, USA}
\affiliation{National Institute of Chemical Physics and Biophysics, 12618 Tallinn, Estonia}

\date{\today}

\begin{abstract}
We study the thermodynamic and high-magnetic-field properties of the magnetic insulator Ba$_5$CuIr$_3$O$_{12}$, which shows no magnetic order down to 2\,K consistent with a spin liquid ground state. While the temperature dependence of the magnetic susceptibility and the specific heat shows only weak antiferromagnetic correlations, we find that the magnetization does not saturate up to a field of 59~Tesla, leading to an apparent contradiction. We demonstrate that the paradox can be resolved, and all of the experimental data can be consistently described within the framework of random singlet states. We demonstrate a generic procedure to derive the exchange coupling distribution $P(J)$ from the magnetization measurements and use it to show that the experimental data is consistent with the power-law form $P(J)\sim J^{-\alpha}$ with $\alpha \approx 0.6 $. Thus, we reveal that high-magnetic-field measurements can be essential to discern quantum spin liquid candidates from disorder dominated states that do not exhibit long-range order.
\end{abstract}

\maketitle
Strong quantum fluctuations in insulating magnetic compounds can give rise to quantum spin liquid (QSL) ground states, where the interaction-driven ordering tendencies are thwarted completely. Devoid of long-range order, QSLs lie beyond the Landau symmetry-based classification, and are characterized instead by their unconventional entanglement properties and the presence of exotic fractionalized excitations \cite{Savary2016,Zhou2017}. 
However, identifying the elusive QSL behavior in real materials has proven to be a formidable task \cite{Norman2016,Zhou2017,Wen2019}. The search for QSL candidate materials represents a major challenge of modern condensed matter physics.

Disorder is one of the major hindrances
to identify QSL materials~\cite{Helton2010,Kimchi2018nat,Choi2019}, as it can drive the formation of random singlet states (RSS) \cite{Kimchi2018prx} or disordered stripe states \cite{Zhu2017} instead of a QSL.
Importantly, this includes single-crystal samples due to intrinsic disorder \cite{Li2015,Li2019}.
A convenient reference point can be found in one-dimensional (1D) systems, where the quantum fluctuations are dominant \cite{Giamarchi2003} and the effect of disorder was clarified some time ago \cite{Ma1979,Hirsch1980}. In 1D it converts the spin liquid ground state into a RSS, where the effective exchange coupling follows a broad probability distribution that has a universal form \cite{Fisher1994} at low energies. In 2D and 3D, on the contrary, the fate of disordered spin systems is still an open question. While a random singlet state with a power-law distribution has been conjectured~\cite{Bhatt1982}, the true ground state of such systems is still under debate and might not be universal \cite{Motrunich2000,Lin2003,Kawamura2014}. In particular, enhanced suppression of QSL states by disorder has been found in model calculations \cite{Kawamura2014,Kawamura2018}. However, mechanisms for the stabilization of QSL states by disorder have also been proposed \cite{Wu2019}. Additionally, a strong spin-orbit coupling (SOC) is an important ingredient in many QSL candidates. While its effects on clean QSLs have been studied \cite{Savary2016,Zhou2017} and particularly emphasized for the so-called Kitaev materials \cite{Knolle2019,Takagi2019}, the interplay of SOC with disorder still remains to be understood.
Thus, careful studies on the role of disorder and SOC in materials showing QSL-like behavior (i.e., no ordering or glassiness down to the lowest temperatures) are of the utmost importance
to confirm, or rule out, the QSL state.

\begin{figure}[t]
\includegraphics[width=0.5\textwidth]{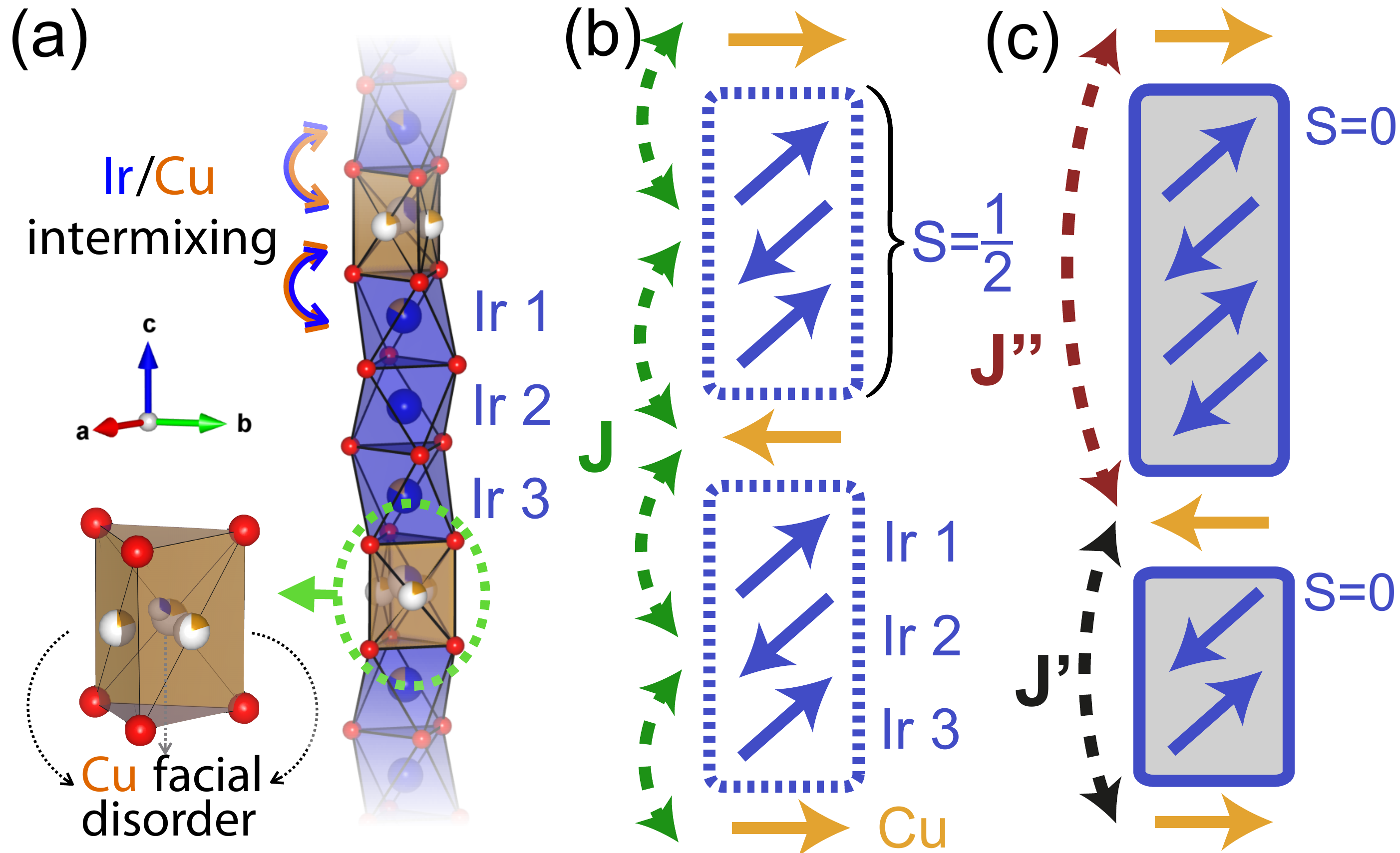}
\caption{The depiction of intrinsic disorder in chains of Cu and Ir in the \BCIO lattice structure.
(a) Cu-Ir chains composed of Ir$^{4+}$ trimers and Cu$^{2+}$ ions (Ba ions fill the space between the chains \cite{Mai2018}). Disorder occurs either due to Cu-Ir site mixing or  due to Cu being displaced from the prism center \cite{Blake1998,Blake1999}.
(b) Spin degrees of freedom in a chain segment, here Ir trimers form effective $J=1/2$ moments that interact with the Cu$^{2+}$ spins.
(c) An example of disorder in the position of Cu and Ir leading to exchange disorder. Interchanging the Cu and Ir sites leads to Ir clusters forming low-spin states. The Cu spins interact with each other through perturbatively generated $J'$ and $J''$ resulting in disorder in the effective magnetic exchange couplings.
}
\label{fig:cartoon}
\end{figure}

In this Rapid Communication, we study the magnetic and thermodynamic properties of the insulating iridate Ba$_5$CuIr$_3$O$_{12}$, which features a quasi-1D arrangement of alternating Cu$^{2+}$ ions and Ir$^{4+}$ trimers \cite{Blake1998,Blake1999} (see Fig.\,\ref{fig:cartoon}). This iridate is of particular interest for the following reasons. First, previous studies \cite{Blake1998} have shown that no magnetic ordering occurs in Ba$_5$CuIr$_3$O$_{12}$ down to 4\,K despite a Curie temperature of $-98$\,K, which suggests a possible QSL ground state. Moreover, a related compound ${\mathrm{Ba}}_{4}{\mathrm{Nb}}{\mathrm{Ir}}_{3}{\mathrm{O}}_{12}$ has recently been proposed to  be a QSL candidate material ~\cite{Nguyen2019}. Second, the nature of the Ir magnetic moments in this system is quite peculiar. The $5d$ Ir ions have a strong spin-orbit coupling and form face-sharing Ir$^{4+}$ trimers, which renders the usual local $J_\mathrm{eff}$=1/2 moment picture \cite{Kim2008,Kim2009} inapplicable due to enhanced covalency. Instead, molecular orbitals at each Ir trimer are expected to form \cite{Streltsov2017,Mai2018,Nguyen2019}. Finally, the material contains intrinsic disorder due to site mixing between Cu and Ir, as well as  Cu displacement from the prism center \cite{Blake1998,Blake1999} [see Fig. \ref{fig:cartoon} (a)]. The former can  lead to randomness (i.e. disorder) in the exchange couplings. A particular scenario is shown in Fig. \ref{fig:cartoon} (c), where interchanging  Cu and Ir within a unit cell transforms two Ir trimers into a dimer and tetramer with a possible $S=0$ ground state. As a result,  the remaining Cu spins interact by means of perturbatively generated exchange couplings, that are different from the initial non-random value. All of the above makes Ba$_5$CuIr$_3$O$_{12}$ a well-suited candidate to explore the interplay of QSL physics with intrinsic disorder and strong spin-orbit coupling.

We have performed magnetic susceptibility, specific heat, and high-field magnetization measurements. We demonstrate that these data combined point unambiguously to Ba$_5$CuIr$_3$O$_{12}$ being in a random singlet state with a power-law distribution of exchange couplings, and thus ruling out QSL behavior. As such, we show how the high field magnetization measurements are essential to reveal and characterize a RSS in materials that otherwise show  QSL-like behavior.

{\it Experimental techniques.} We have grown single crystals of \BCIO using the flux method. The crystal structure and orientation were confirmed by x-ray diffraction and Laue measurements \footnote{See Supplemental Material below for the details of crystal growth procedure, calibration of the high-field measurements, random singlet model and the fitting procedure as well as magnetic anisotropy and zero-field cooled measurements, which includes Refs. \cite{Blake1998,Blake1999,khaliullin.2006,Rau2016,Bertinshaw2019,Modic2014,ishizuka.2017,Mai2018,Nguyen2019}.}. The magnetic susceptibility was measured using a superconducting
quantum interference device (SQUID) magnetometer (Quantum Design) in an applied field of 0.1\,T on warming after zero-field cooling to 1.8\,K. The specific heat of \BCIO single crystals was measured using a Physical Property Measurement System (Quantum Design Dynacool). The high-field magnetization measurements have been performed at 2\,K in pulsed magnetic fields up to 59\,T \cite{Note1} using the facilities at the Dresden High Field Magnetic Field Laboratory, described in Refs. \cite{Zherlitsyn2006,Wosnitza2007,Skourski2011}.

\begin{figure}
\includegraphics[width=0.50\textwidth]{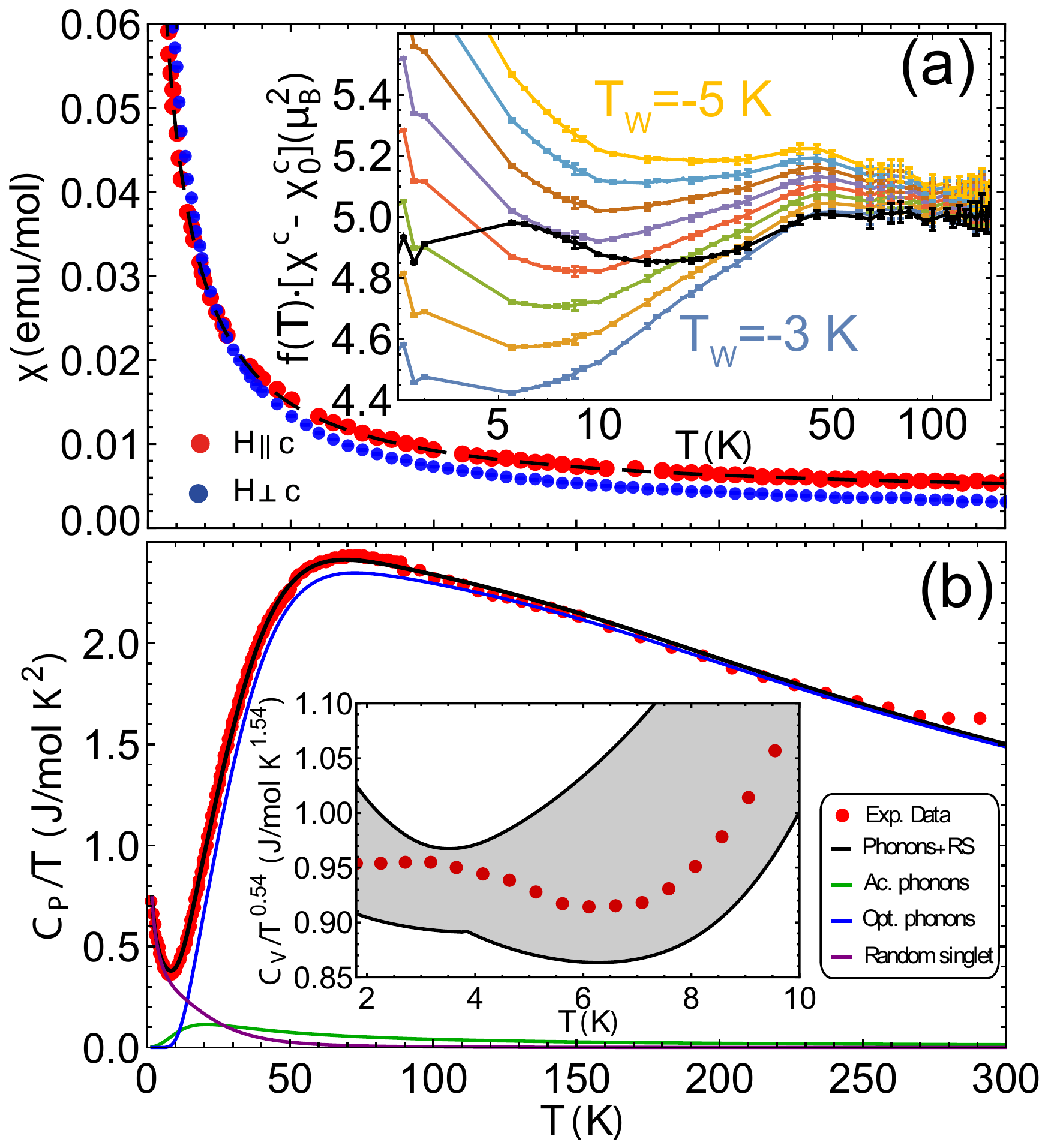}
\caption{Temperature ($T$) dependencies of the magnetic susceptibility ($\chi$) and the specific heat ($C_P$).
(a) The magnetic susceptibility, data in red ($H\parallel c$) and blue ($H\perp c$).
The black dashed line is a fit for $H\parallel c$ with the random singlet model $\chi_{RS}=\partial M_{RS}/\partial H$, see Eq. \eqref{eq:Mrsm}. Inset: $(\chi^c-\chi^c_0)$ multiplied by a function $f(T)$. For the colored points we take $f(T)=3(T-T_W)$ for several values of $T_W$ between -3\,K and -5\,K, for $H\parallel c$ demonstrating the nonlinearity of the low temperature dependence.
Black points are the RSS contribution $f(T)=\chi_{RS}(T)/\mu_{eff}^2$. Lines are guide to the eye. At high temperatures all curves converge to $\mu_{eff}^2$.
(b) Specific heat divided by temperature. The black line is a fit to the combination of the random singlet model in Eq. \eqref{eq:Crsm} and a simplified model for phonons (see text). Inset: The specific heat divided by $T^{0.54}$; the gray band shows the confidence interval of the fit.
}
\label{fig:temp}
\end{figure}

{\it Magnetic susceptibility.}
In Fig.~\ref{fig:temp}(a) we show the temperature dependence of the magnetic susceptibility for fields along the $c$ axis $\chi^{c}(T)$ or in the $a-b$ plane $\chi^{ab}(T)$. Both $\chi^{c}(T)$ and $\chi^{ab}(T)$ show a featureless monotonic increase towards low temperatures and a weak anisotropy \cite{Note1}. At high temperatures, a constant contribution $\chi_0$ in addition to the Curie behavior can be identified, which is attributed to Van Vleck paramagnetism. The effective moment that is obtained  from the Curie law fit is $\mu_\mathrm{eff}=2.2\mu_B$, which is close to the value that is expected from one Cu$^{2+}$ moment ($\mu_\mathrm{eff}^\mathrm{Cu}=1.9\mu_B$) and one Ir trimer ($\mu_\mathrm{eff}^\mathrm{Ir-tr} = 0.8\mu_B$) \cite{Nguyen2019} that yield $\sqrt{(\mu_\mathrm{eff}^\mathrm{Cu})^2+(\mu_\mathrm{eff}^\mathrm{Ir-tr})^2} \approx2.06\mu_B$.

In an earlier study on polycrystalline samples \cite{Blake1998}, the susceptibility was analyzed using a modified Curie-Weiss model for temperatures between 150\,K and 300\,K. Using $\chi(T)=C/(T-T_{W})+\chi_0$
resulted in a large negative Weiss temperature $T_W =-98$\,K. Analyzing our data, we have come to the conclusion that the Curie-Weiss model does not provide an adequate description. First, if the analysis is restricted to high temperatures, large uncertainties in the value of $T_W$ result \cite{Note1}. Second, at low temperatures $(\chi-\chi_0)^{-1}$ is not linear as would have been expected from the Curie-Weiss form. We demonstrate this in the inset of Fig. \ref{fig:temp} (a) by showing $(T-T_W)(\chi^c-\chi^c_0)^{-1}$ for a range of $T_W$ from $-3$\,K to $-5$\,K. Additionally, one can see that larger or smaller values of $T_W$ would lead to even larger deviations, suggesting weak antiferromagnetic (AFM) correlations.


This is further corroborated by noting that even in the absence of order, anomalies in $\chi(T)$ are expected to arise at a temperature corresponding to the interaction scale in 1D antiferromagnetic chains \cite{Bonner1964,Eggert1994}, spin glasses \cite{Binder1986,Mydosh2015}, and spin liquids with AFM interactions \cite{Yoshitake2016}. The absence of such features in Fig. \ref{fig:temp} (a) implies that the relevant interaction scale is lower than 2\,K. We have also confirmed the absence of glassy behavior above 2\,K by performing low-field (100 Oe) field-cooled/zero-field-cooled (FC/ZFC) susceptibility measurements \cite{Note1}.

{\it Specific heat.}
In Fig. \ref{fig:temp}(b) we show the temperature dependence of the specific heat $C_P(T)$. The high-temperature behavior of $C_P(T)/T$ is dominated by the phonon contribution, which freezes out as the temperature is lowered. Thus, the dramatic upturn that is observed below $\sim 10$\,K must be of magnetic origin. As no Schottky-like peak is observed down to 2\,K, the energy scale associated with these magnetic excitations should be below 2\,K. This is  consistent with the weak AFM correlations conjectured above on the basis of the $\chi(T)$ measurements.

\begin{figure}
\includegraphics[width=0.50\textwidth]{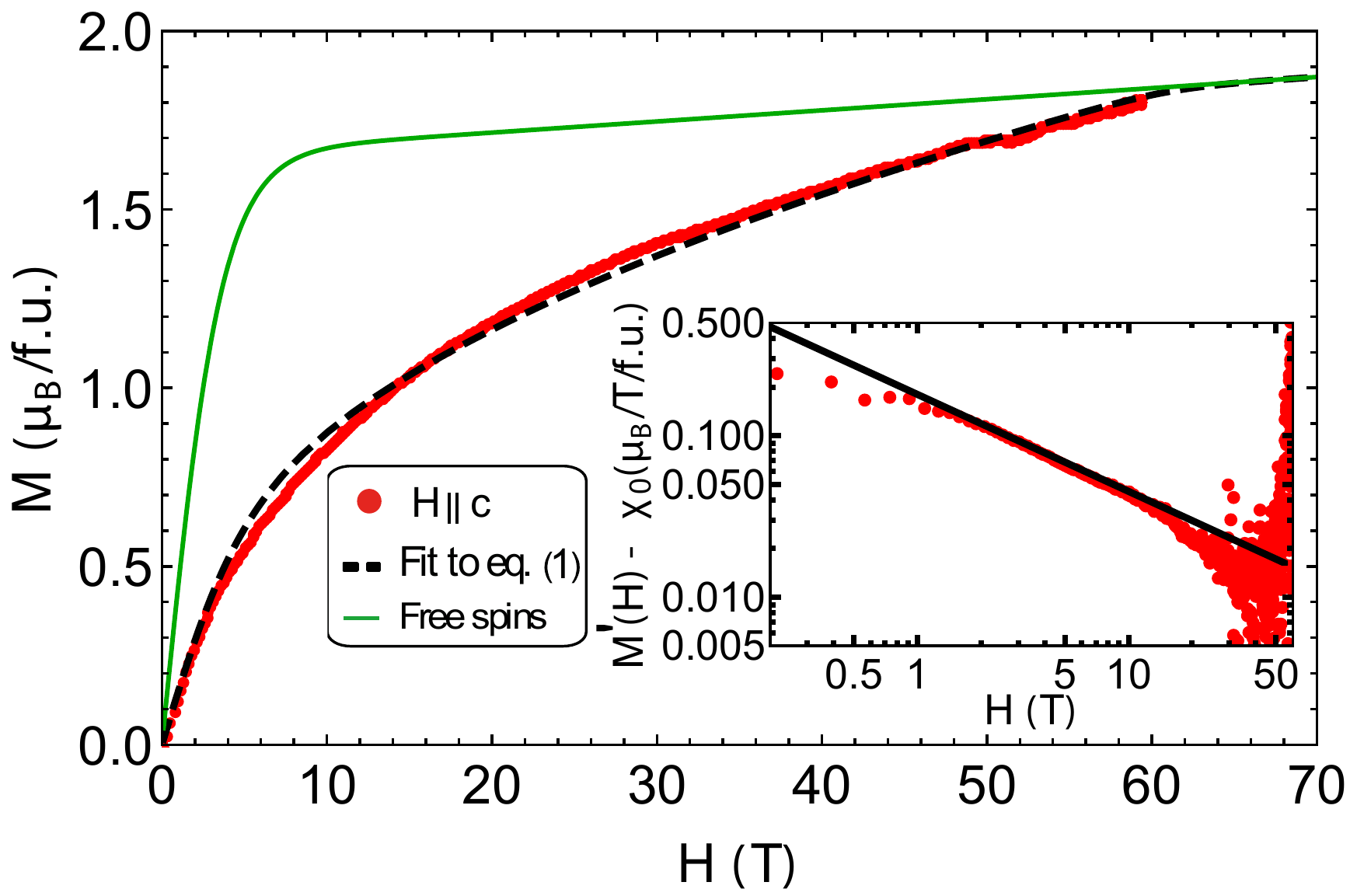}
\caption{The magnetic field dependence of the magnetization at $T=2$ K for the  field along the $c$-axis direction. The weak kink near 50\,T results from the noise of the equipment. The black dashed line is a fit with the random singlet model, Eq. \eqref{eq:Mrsm}, using the parameters given in Table \ref{table:Fit}. The green line represents the magnetization of an $S=1/2$ paramagnet. The Van Vleck contribution $H\chi_0$ has been added to both. Inset: Log-Log plot of  $d M(H)/dH-\chi_0$ for the field along the $c$-axis, and the black line is a power-law fit $0.18 H^{-0.6}$. Data for $H\parallel ab$ are not shown due to calibration issues \cite{Note1}.
}
\label{fig:Mag}
\end{figure}

\begin{figure}[t]
\includegraphics[width=0.4\textwidth]{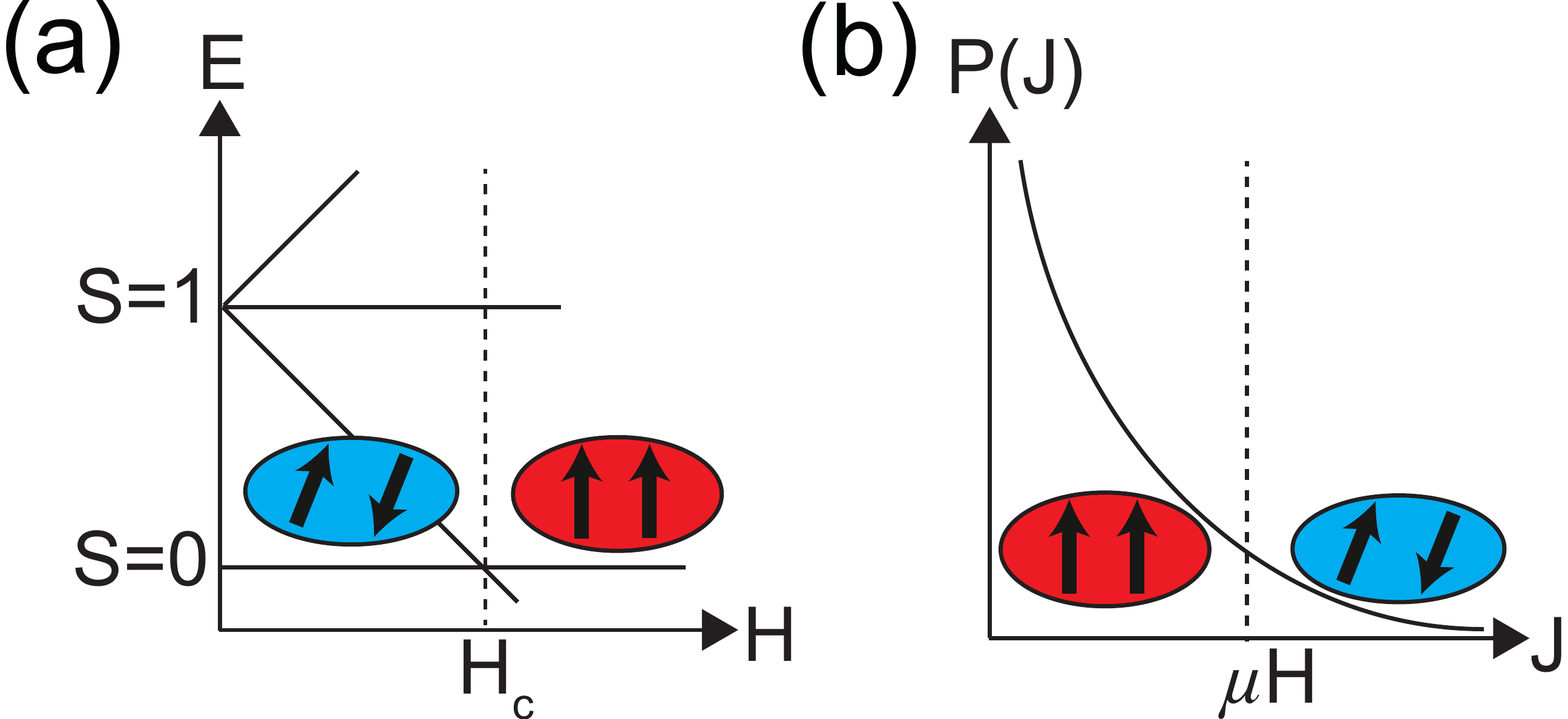}
\caption{ (a) The energy levels and the ground state of an isolated singlet. The triplet ($S=1$) of excited states at $H=0$ is split in the field, and a change of the ground state occurs at $H_c(J)$, from singlet ($S=0$) to fully polarized ($S=1$).
(b) The random singlet distribution in a magnetic field. Singlets with $J<\mu H$ are broken by the field and are fully polarized, while the ones with $J>\mu H$ remain in the singlet state, leading to a non-saturating magnetization.
}
\label{fig:cartoon2}
\end{figure}

{\it High-field magnetization.} Surprisingly, the field dependence of the magnetization $M(H)$ is in stark contrast with the expectation from weak AFM correlations, see Fig.~\ref{fig:Mag}. Namely, $M(H)$ shows a monotonic increase without saturation up to the highest fields measured, 59\,T. To illustrate this, we show in Fig. \ref{fig:Mag} (green line) the $M(H)$ that is expected for a system of two free $S=1/2$ spins per unit cell, with an effective moment $\mu_\mathrm{eff}/\sqrt{2}$ each, and taking the Van Vleck contribution $M_{VV}=\chi_0 H$ into account. 
One can see that within such a model the magnetization would have saturated well below 59\,T, implying that the magnetic interactions in \BCIO must be rather strong.
One can estimate the scale of the interactions assuming the $S=1/2$ moments mentioned above to form singlets with an isotropic exchange energy $J$.  The magnetization would then saturate when the Zeeman energy
$E_Z=H g S = H \mu_\mathrm{eff}\sqrt{S/(S+1)}$
for the triplet excitation reaches $J$, see Fig. \ref{fig:cartoon2}.
As the saturation field is at least larger than 59\,T, we estimate $J\gtrsim 70$\,K. On the contrary, the energy scales we have derived above from the susceptibility and specific heat measurements are below 2\,K. In addition, in systems with AFM interactions the shape of the magnetization curve as a function of $H$ is usually convex \cite{Bonner1964,Griffiths1964,Kashurnikov1999,Shimokawa2015} at low temperatures, while the $M(H)$ curve shown in Fig. \ref{fig:Mag} is clearly concave, further making the interpretation of the high-field magnetization in terms of a strong AFM exchange interaction problematic.

{\it Random singlet state.} We will now show that the conflict between the energy scales that we have seen in low- and high-field measurements can be resolved by assuming a {\it distribution} of energy scales in the system in the framework of a RSS. The exchange disorder driving the RSS can result from the intrinsic positional disorder between Cu and Ir observed in x-ray \cite{Blake1998} and neutron scattering \cite{Blake1999} experiments, as discussed above and illustrated for a particular scenario in Fig. \ref{fig:cartoon} (c). Other possible scenarios would involve nonstoichometric compositions within one unit cell, e.g., simply substituting one Cu for Ir.


Given the small magnetic anisotropy observed in $\chi(T)$ [see Fig. \ref{fig:temp} (a)], we consider an
ensemble of singlets formed by two effective $S=1/2$ moments with a total magnetic moment $\mu$, and with an isotropic random exchange coupling $J$ that is drawn from the distribution $P(J)$. The magnetization of the whole system is then an average of the magnetization of each isolated singlet, and is given by
\begin{equation}
M_{RS}(H) =\int_0^{\infty} dJ P(J)\frac{2 \mu \sinh(\beta\mu H)}{2 \cosh(\beta\mu H)+1+e^{\beta J}},
\label{eq:Mrsm}
\end{equation}
where $\beta=1/(k_B T)$. We account for the Van Vleck contribution as before for free spins, i.e., $M(H) =M_{RS}(H) +H \chi_0 $. Qualitatively, Eq. \eqref{eq:Mrsm} allows a coexistence of almost free spins that can yield a diverging susceptibility towards $T=0$ and strongly bound singlets from the high-$J$ tail of the distribution that require the applied field to be above a threshold value for the magnetization to saturate (see Fig. \ref{fig:cartoon2}). Importantly, in the limit $T\ll \mu H$ one obtains from Eq.~\eqref{eq:Mrsm} that $M'(H)\approx \chi_0+\mu^2 P(\mu H)$, allowing one to extract the functional form of the distribution $P(J)$ directly from the experimental data. We find that $P(\mu H)$  follows the power-law form $P(\mu H)\sim H^{-0.6}$ for fields between 1 and 15\,T (see Fig. \ref{fig:Mag}, inset).

Let us now discuss the specific heat. Similarly to the magnetization, the contribution of the RSS is an average over specific heats of individual singlets
\begin{equation}
C_{RS}(T)= k_B \int_0^{\infty} dJ P(J) \frac{J^2}{T^2} \frac{3e^{-J/T}}{(1+3 e^{-J/T})^2}.
\label{eq:Crsm}
\end{equation}
For $P(J)\sim J^{-\alpha}$ it follows that at low temperatures $C_{RS}\sim T^{(1-\alpha)}$. Indeed, we find that below about 4\,K, $C_P\sim T^{0.54}$ [see Fig. \ref{fig:temp} (b), inset], that suggests the power-law exponent to be $0.46$. The discrepancy of this value with the one obtained from the high-field magnetization can be attributed to $P(J)$ having a slightly different form for low and moderate $J$, as the specific heat \eqref{eq:Crsm} is most sensitive to $P(J)$ below $J\approx 4\,$K, while the power-law in the magnetization is extracted for larger values of $J$. Nonetheless, the discrepancy between the power-law exponents is not too large.

Hence, we have attempted to fit the data from each measurement with a single form of $P(J)=\theta(J_0-J) J^{-\alpha}$, where a cutoff scale $J_0$ has been introduced to ensure proper normalization. The results of the fits are given in Table \ref{table:Fit}. The parameter $\mu$ in Eq. \eqref{eq:Mrsm} is related to $\mu_{\mathrm{eff}}$
at high temperatures as $\mu=\sqrt{2/3}\;\mu_{\mathrm{eff}}$. Additionally, to describe the specific heat at all temperatures, we have modeled the phonon contribution of specific heat with a combination of Debye and Einstein phonons \cite{Note1}, i.e. $C_P(T)=C_{RS}(T)+C_{\mathrm{phon}}(T)$.

\begin{table}
\caption{\label{table:Fit}
The power-law exponents $\alpha$ and the cutoff scale $J_0$ are obtained by fitting the magnetic susceptibility, magnetization, and heat capacity data with the corresponding confidence intervals. The resulting fits are plotted in Figs. \ref{fig:temp} and \ref{fig:Mag}.}
\begin{ruledtabular}
\begin{tabular}{lccc}
Measured
Quantity&$\alpha$&$J_0$\,(K)\\
\hline
$\chi^c(T)$&$0.62\pm0.02$&$36\pm1$\\
$\chi^{ab}(T)$&$0.66\pm0.01$&$16.3\pm0.4$\\
$M_c(H)$&$0.64\pm0.01$&$67.8\pm0.4$\\
$C_p(T)$&$0.55\pm0.05$&$95\pm5$\\
\end{tabular}
\end{ruledtabular}
\end{table}

The resulting fits to  the data are excellent as shown in Figs. \ref{fig:temp} and \ref{fig:Mag}. Importantly, the qualitative features of all three measurements are well captured: the susceptibility increasing nonlinearly at low-$T$
[see the inset of Fig. \ref{fig:temp} (a)],
the upturn in the specific heat at low-$T$ where $C_{RS}$ dominates, and the non-saturating concave high-field magnetization. Moreover,  the resulting power-law exponents obtained from fits across different experiments  agree well with each other (see Table \ref{table:Fit}). The cutoff scale $J_0$, on the contrary, shows significant variations. 
This can be partially attributed to the deviations of $P(J)$ from the power-law form at the lowest and highest values of $J$ (as is seen in Fig. \ref{fig:Mag}), as different quantities are most sensitive to different ranges of $J$ values. Additionally, it can be shown that this parameter depends on the way the cutoff is implemented - e.g., implementation of a soft cutoff affects the value of $J_0$ \cite{Note1}. Thus, we argue that the variations of $J_0$ reflect the approximate character of the form of $P(J)$ we use, which is nonetheless sufficient for the qualitative description of the data.

As has been mentioned above, the distribution parameters may vary between the low and intermediate energy scales. The agreement of the power-law exponents in Table \ref{table:Fit} with the one obtained from magnetization between 1 and 15\,T suggests that these values do not concern the distribution at very low energies. Instead, we have established the presence of random singlet excitations with a unique power-law form in the intermediate energy range.

{\it Summary.} By combining low- and high- magnetic field measurements we have established that \BCIO at low temperatures is well described as a random singlet state. We have shown that a non-saturating high-field magnetization allows one to rule out a QSL scenario and quantitatively extract the exchange coupling distribution of the random singlet state $P(J)\sim J^{-0.6}$ at intermediate energies. We find the extracted power-law  distribution is consistent across the magnetization, susceptibility, and specific heat measurements. Thus, we establish that a combination of high-field measurements with more conventional techniques allows one to study the role of disorder in QSL candidate materials as well as characterize strongly disordered ground states.


{\it Acknowledgments. } Crystal growth was supported by the National Research Foundation of Korea, Ministry of Science and ICT (No. 2016K1A4A4A01922028), and magnetic measurement was supported by the NSF under Grant No. DMR-1629059. We acknowledge the support of HLD at HZDR, member of the European Magnetic Field Laboratory (EMFL). The work was supported by the DFG through SFB 1143. The spectroscopic characterization at Rutgers (M.Y. and G.B.) was supported by NSF Grant No. DMR-1709161. J. H. P. was supported by Grant No. 2018058 from the United States-Israel Binational Science Foundation (BSF), Jerusalem, Israel and acknowledges the Aspen Center for Physics where some of this work was performed, which is supported by National Science Foundation Grant No. PHY- 1607611. P.A.V. acknowledges the support by the Rutgers University Center for Materials Theory Postdoctoral fellowship.

\newpage

\clearpage
\onecolumngrid
\appendix
\renewcommand{\thefigure}{S\arabic{figure}}
\addtocounter{equation}{-2}
\addtocounter{figure}{-4}
\addtocounter{table}{-1}
\renewcommand{\theequation}{S\arabic{equation}}
\renewcommand{\thetable}{S\arabic{table}}

\begin{center}
	\textbf{\Large
		Supplemental Material for: \\Random singlet state in Ba$_5$CuIr$_3$O$_{12}$ single crystals}
\end{center}

\section{Single Crystal Growth}
For the \BCIO single crystal growth, we first prepared polycrystalline material by the solid-state reaction method: a stoichiometric composition of BaCO$_3$, IrO$_2$ and CuO powders (Alfa Aeser) were grounded, and the pelletized powder was sintered in the air at 1000 -- 1200\,C with intermediate grindings. Then \BCIO single crystals were grown by flux method using K$_2$CO$_3$ with added BaCO$_3$. The \BCIO polycrystalline powder were mixed with the flux of K$_2$CO$_3$ and BaCO$_3$, and then were melted in alumina crucible at 1050\,$^\circ$C and slowly cooled down to room temperature at a cooling rate 2 $^\circ$C/hr. The crystals were washed with hot water to separate from the flux.

\begin{centering}
	\begin{figure}[h!]
		\includegraphics[width=0.9\textwidth]{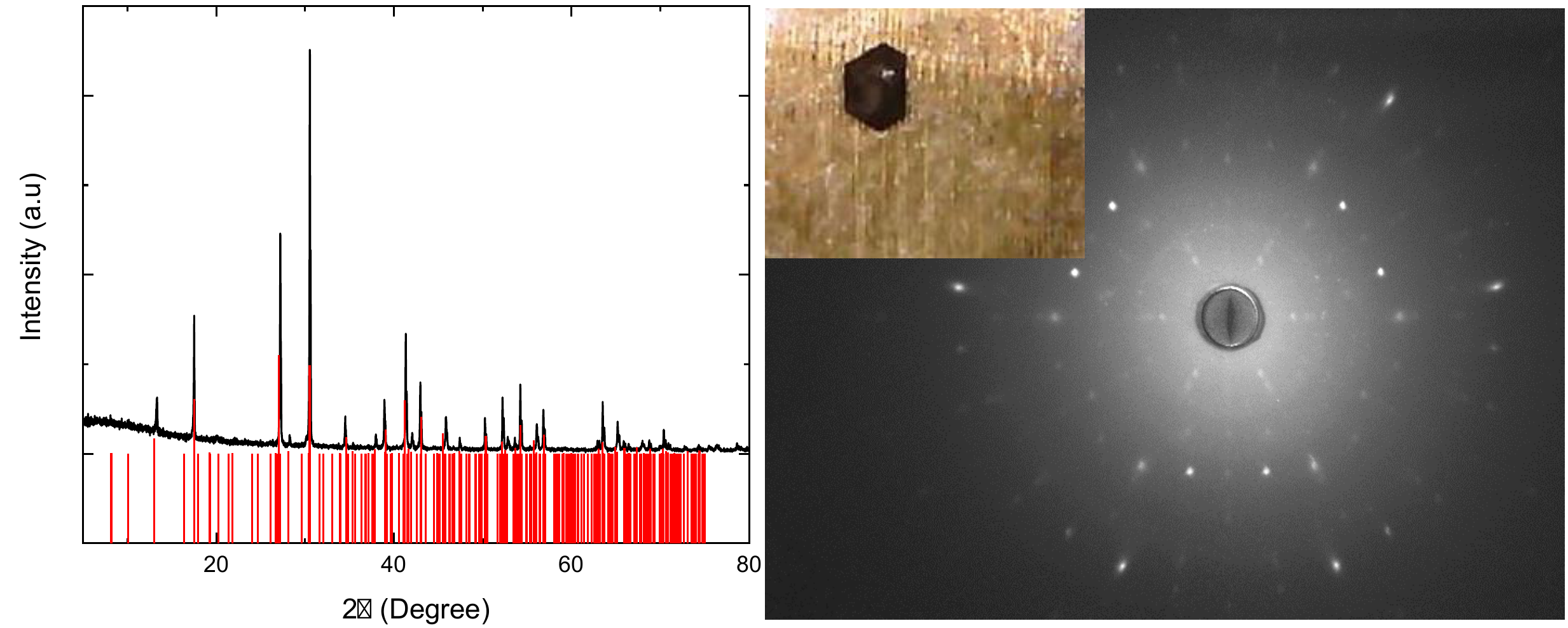}
		\caption{Left panel: XRD measurement data (black) and intensity pattern expected form \BCIO (red lines). Right panel: Laue diffraction data confirming the threefold symmetry and single-crystal nature of the sample. Inset: image of one of the samples on the Laue holder.
		}
		\label{sup:fig:sample}
	\end{figure}
\end{centering}

XRD measurements of crushed single-crystal powder confirm the right phase (see Fig. \ref{sup:fig:sample}); the resulting lattice parameters are a=10.1361\AA, c=21.3561\AA. While a agrees to within 0.1$\%$ with the previous measurements \cite{Blake1998,Blake1999} c is shorter by about 1.4$\%$ than the previous reported values \cite{Blake1998,Blake1999}. We attribute this discrepancy as well as imperfect match to the predicted intensities to the possible strains and imperfections emerging during crushing. Additionally, we have performed Laue diffraction measurements; the resulting pattern for (001) direction is presented in Fig. \ref{sup:fig:sample}, right panel. It confirms the single-crystal character of our samples and the threefold pattern symmetry is consistent with the $P3c1$ space group of \BCIO.

%
%
%
%
%
%

\section{Model of the random singlet state}

The partition function of a singlet with antiferromagnetic (AFM) interaction $J$ in magnetic field $H$ is (with the singlet state energy set to zero)
\begin{equation}
Z_s=1+ e^{-\beta J}[2 \cosh(\beta\mu H)+1].
\label{sup:eq:partition}
\end{equation}
And the corresponding magnetization is
\begin{equation}
M_s= -\frac{\partial F}{\partial H}=\mu\frac{2 \sinh(\beta\mu H)}{2 \cosh(\beta\mu H)+1+e^{\beta J}}.
\label{sup:eq:magnetization}
\end{equation}
The magnetization of the whole system is an average of $M_s$ over a distribution $P(J)$:
\begin{equation}
M =\int_0^{\infty} dJ P(J) M_s= \int_0^{\infty} dJ P(J) \mu\frac{2 \sinh(\beta\mu H)}{2 \cosh(\beta\mu H)+1+e^{\beta J}}.
\label{sup:eq:magnetizationWhole}
\end{equation}
On the one hand, if $T\ll\mu H\,\&\,J$, $M_s$ is equal to $\mu$ for $\mu H>J$ and to $0$ otherwise. In this limit, Eq.(\ref{sup:eq:magnetizationWhole}) simplifies to
\begin{equation}
M = \mu \int_0^{\mu H} dJ P(J) +O[TP(\mu H)].
\label{sup:eq:magnetizationSmallT}
\end{equation}
The derivative of M with respect to $H$ is then
\begin{equation}
M'(H) = \mu^2 P(\mu H).
\label{sup:eq:mat0}
\end{equation}
On the other hand, if $T\gg\mu H\,\&\,J_{0}$, we have
\begin{equation}
M = \int_0^{\infty} dJ P(J) \mu\frac{2 \beta\mu H}{4}=\frac{\mu^2 H}{2T},
\label{sup:eq:mathigh1}
\end{equation}
and
\begin{equation}
M'(H) = \frac{\mu^2}{2T}.
\label{sup:eq:mathigh2}
\end{equation}
Eq.(\ref{sup:eq:mathigh2}) is simply the Curie susceptibility $\frac{(g\mu_B)^2S(S+1)}{3T}$ of two S=1/2 spins with g-factor $\mu/\mu_B$. We can also calculate the zero-field specific heat from Eq.(\ref{sup:eq:partition}):
\begin{equation}
C_{RS}= - T k_B \frac{\partial^2 F} {\partial T^2} = \int_0^{\infty} dJ P(J) \frac{J^2}{T^2} \frac{3e^{-J/T}}{(1+3 e^{-J/T})^2}.
\label{sup:eq:cv}
\end{equation}

Using Eq.(\ref{sup:eq:mat0}) one can extract the distribution $P(J)$ from the low-temperature field dependence of magnetization. Using (\ref{sup:eq:mathigh2}) the value of the moment $\mu$ can be extracted from the high-temperature susceptibility value.

\section{Magnetic anisotropy}
\begin{centering}
	\begin{figure}[h!]
		\includegraphics[width=0.9\textwidth]{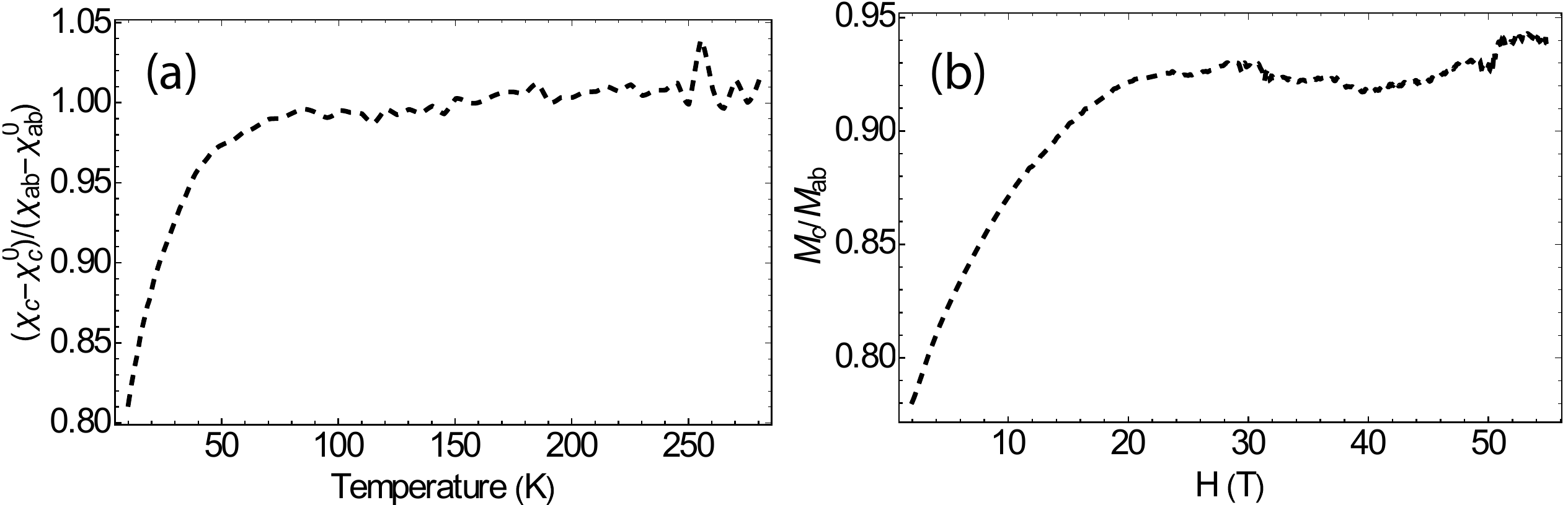}
		\caption{Anisotropy of (a) magnetic susceptibility (b) magnetization, as a function of temperature and magnetic field, respectively.
		}
		\label{sup:fig:anis}
	\end{figure}
\end{centering}

In the main text we have concentrated on the data for $H\parallel c$ and generally disregarded the magnetic anisotropy. While it is indeed weak, we provide here additional information regarding this anisotropy and its dependence on the applied field and temperature. In Fig. \ref{sup:fig:anis} we show the anisotropy of  (a) magnetic susceptibility (b) magnetization, as a function of temperature and magnetic field, respectively. Note that the van Vleck contribution, subtracted form the data in the figure, is also anisotropic: $\chi_0^c= 0.0033$ emu/mol and $\chi_0^ab=0.0012$ emu/mol. To account for the discrepancies in the temeprature/field values of the two measurements we use the numerically interpolated values. At large fields or temperatures the anisotropy is seen to decrease. This can be attributed to the anisotropic exchange interactions, characteristic of systems with strong spin-orbit coupling \cite{khaliullin.2006} and Ir-based ones in particular \cite{Rau2016,Bertinshaw2019}. Indeed, temperature-dependent anisotropy has been observed in Ir-based systems \cite{Modic2014} and is expected theoretically \cite{ishizuka.2017}.

\section{FC/ZFC measurements and glassiness}
\begin{centering}
	\begin{figure}[h!]
		\includegraphics[width=0.9\textwidth]{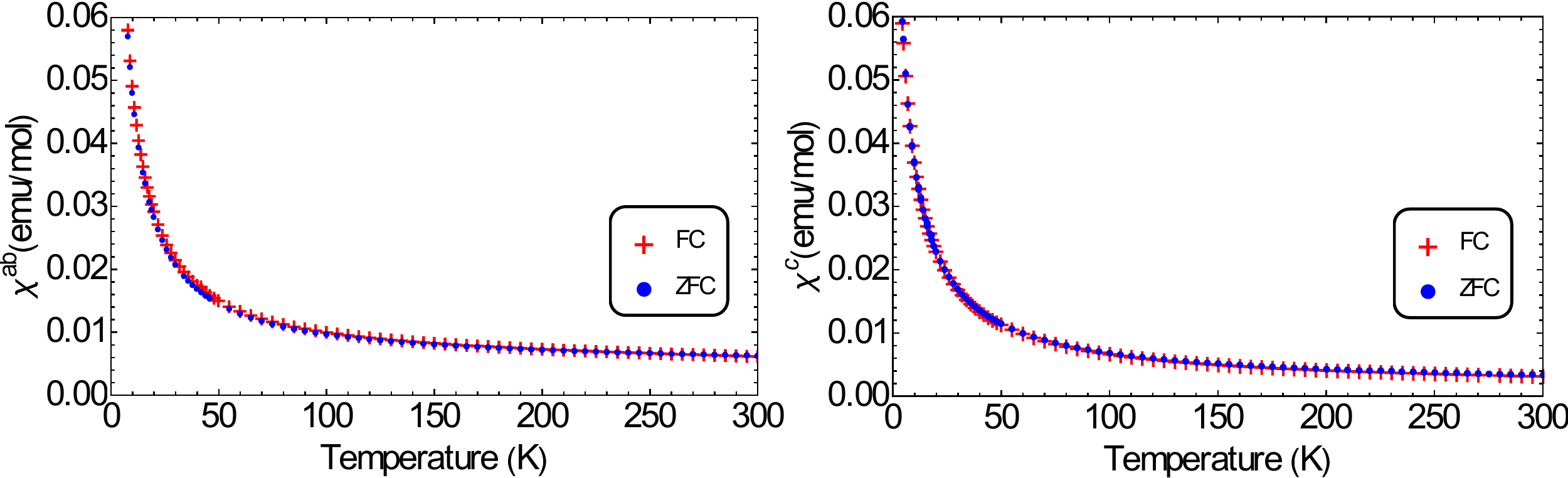}
		\caption{Comparison of FC and ZFC susceptibility data.
		}
		\label{sup:fig:fczfc}
	\end{figure}
\end{centering}

To confirm the absence of glassy behavior in Ba$_5$CuIr$_3$O$_{12}$ we have performed ZFC/FC magnetic susceptibility measurements in field of 100 Oe using a SQUID magnetometer (Quantum Design), with the results presented in Fig. \ref{sup:fig:fczfc}. We find no signatures of spin freezing or history dependence. The relative differences between the FC and ZFC data are below $2\%$ for both field orientations (for points measured at the same temperature) and can be attributed to the equipment noise (especially taking into account the relatively large background signal, which has been observed to be roughly 10 times larger at this field value then that of an optimally working SQUID).

\section{Details of fitting the data}
\subsection{General remarks and choice of random coupling distribution}
In main text we have used distribution $P(J)=\theta(J_0-J) J^{-\alpha}$ for the exchange interaction in the random singlet state, motivated by the magnetization data in Fig. 3, inset. The $\theta(J_0-J)$ factor is necessary as the integral over the distribution otherwise diverges at large $J$. While this allows us to satisfactorily describe the data qualitatively, this distribution form is only approximate and thus the values of the parameters extracted from the fits can be model-dependent. Here we show that using a different different way to implement the cutoff may strongly affect the value of $J_0$, but not $\alpha$. Namely, we attempt to fit the experimental data with the distribution $P(J,\sigma)=\frac{ J^{-\alpha}}{e^{(J-J_0)/\sigma}+1}$, where $\sigma\to0$ results in the distribution used in the main text. The results of the fits for several values of $\sigma$ are given in Table \ref{table:SupFit}. Indeed, the values of $J_0$ are affected by $\sigma$, especially for the observables yielding lower values of $J_0$, such as the susceptibility $\chi_c(T)$. On the other hand, the power-law exponent $\alpha$ shows a less pronounced dependence on $\sigma$ for all datasets. Additionally, the discrepancies between the fit results for different observables can be explained as follows. While $C_p(T)$ and $\chi(T)$ are primarily sensitive to the low-energy part of the distribution, $M(H)$ probes instead the intermediate-to-high energy part most reliably. Indeed, the behavior of $C_p(T)$ at low T (see Fig. 2 (b) of the main text) suggests $\alpha\approx 0.46$, different from the one extracted from $M(H)$. As we use a distribution having same form at low and intermediate energies, one may expect such discrepancies to appear.	Additionally, we note that phonons provide an important contribution to $C_p(T)$  and thus the parameters of the random-singlet model also may depend on the modeling of phonon contribution (see below).

\begin{table}[h!]
	\caption{\label{table:SupFit}
		The power-law exponents $\alpha$ and the cutoff scale $J_0$ are obtained by fitting the magnetic susceptibility, magnetization, and heat capacity data with the corresponding confidence intervals with a soft cutoff characterized by the width $\sigma$. Fits for $\chi^{ab}$ for $\sigma>10$\,K did not converge to a satisfactorily accuracy.}
	\begin{ruledtabular}
		\begin{tabular}{lccccc}
			Measured
			Quantity, $\sigma$ &$\alpha$ ($\sigma=0$\,(K)) & $\alpha$ ($\sigma=5$\,(K))&$\alpha$ ($\sigma=10$\,(K)) &$\alpha$ ($\sigma=15$\,(K))\\
			\hline
			$\chi^c(T)$&$0.62\pm0.02$&$0.60\pm0.02$&$0.56\pm0.02$&$ 0.5 \pm 0.04$\\
			$\chi^{ab}(T)$&$0.66\pm0.01$&$0.62\pm0.02$&---&---\\
			$M_c(H)$&$0.64\pm0.01$&$0.62\pm0.01$&$0.59\pm0.02$&$0.54\pm0.02$\\
			$C_p(T)$&$0.55\pm0.05$&$0.54\pm0.06$&$0.54\pm0.06$&$0.54\pm0.07$\\
		\end{tabular}
	\end{ruledtabular}
	\begin{ruledtabular}
		\begin{tabular}{lccccc}
			Measured
			Quantity, $\sigma$ 
			& $J_0$\,(K) ($\sigma=0$\,(K))& $J_0$\,(K) ($\sigma=5$\,(K))& $J_0$\,(K) ($\sigma=10$\,(K))& $J_0$\,(K) ($\sigma=15$\,(K))\\
			\hline
			$\chi^c(T)$&$36\pm1$&$34\pm1$&$26\pm2$&$7\pm5$\\
			$\chi^{ab}(T)$&$16.3\pm0.4$&$10\pm1$&---&---\\
			$M_c(H)$&$67.8\pm0.4$&$65.8\pm0.4$&$60.5\pm0.5$&$52.3\pm0.5$\\
			$C_p(T)$&$95\pm5$&$95\pm7$&$96\pm7$&$97\pm8$\\
		\end{tabular}
	\end{ruledtabular}
\end{table}

\subsection{Susceptibility and Curie-Weiss law}
Here we provide some details regarding the deviation of $\chi^{c,ab}(T)$ from Curie-Weiss behavior. Analyzing our data in the same way as in \cite{Blake1998}, we found large uncertainties in $T_W$, e.g. for the interval between 150\,K and 300\,K we get $23\pm15$\,K along the $c$-axis and $-16\pm7$\,K for the in-plane (i.e. $a-b$) direction. Expanding the fitting interval to include lower temperatures (while keeping the upper bound of 300\,K), we find that the uncertainties diminish, and that the value of T$_W$ for the lower bound under 40\,K falls between $-3$\,K and $-5$\,K along the $c$-axis and between $-1$\,K and $-3$\,K for the in-plane direction. These results suggest weak antiferromagnetic (AFM) correlations. 

\begin{figure}[h!]
	\includegraphics[width=0.8\textwidth]{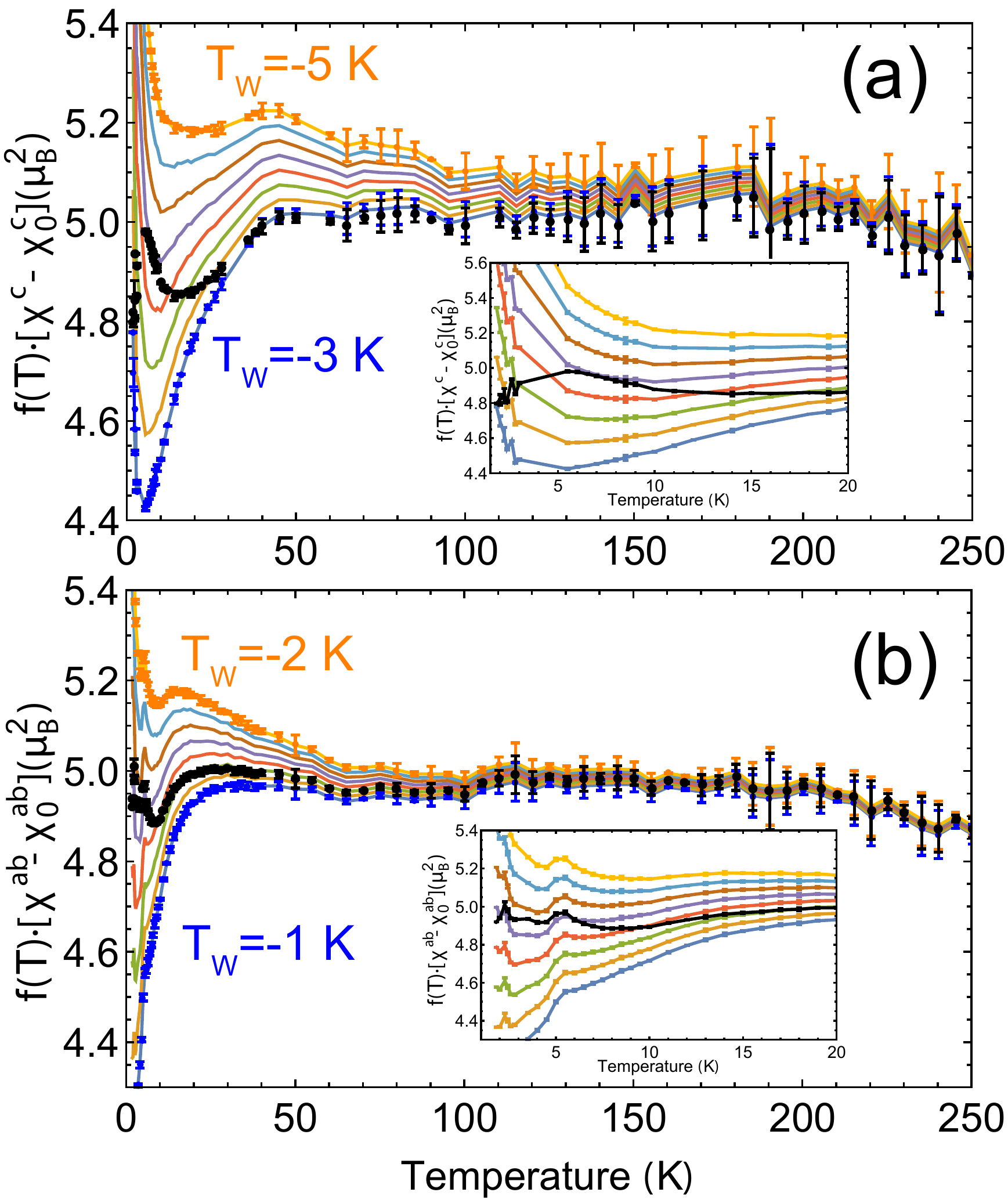}
	\caption{(a) $\chi^c-\chi^c_0$ multiplied by a function $f(T)$. For colored points $f(T)=3(T-T_W)$ for several values of $T_W$ between -3\,K and -5\,K, for $H\parallel c$. Black points: $f(T)=\chi_{RS}(T)/\mu_{eff}^2$. Lines are guide to the eye. At high temperatures all curves converge to $\mu_{eff}^2$. (b) Same for $\chi^{ab}-\chi^{ab}_0$ and values of $T_W$ between -1\,K and -2\,K. Kink at around 2\,K is due to the equipment noise. The units have been converted from $emu\cdot K/mol$ to $\mu_B^2$ with a prefactor $k_B[erg/K]/N_A/\mu_B^2[erg/G]$.
	}
	\label{sup:fig:cwfit}
\end{figure}

In Fig. \ref{sup:fig:cwfit} we present $\chi-\chi_0$ multiplied by a function $f(T)$ for fields in the chain (a) and in-plane (b) directions. Colored lines are for $f(T)=3(T-T_W)$ and black line corresponds to $f(T)=\chi_{RS}(T)/\mu_{eff}^2$. At high temperatures all curves converge to the same value (within error bars). Note that the error bars grow with $T$ as we effectively multiply $\chi$ by $T$ at high temperature, and thus the absolute value of the error bar for is enhanced with it. At low temperatures one can see that the colored curves exhibit significant deviations from the high-temperature value. For some values of $T_W$ there is a pronounced dip, while for others it is the marked enhancement towards low $T$. However, there is no single curve that has neither. In this respect, fits with the RSS model show superior quality (see insets for low temperatures).

\subsection{Specific heat}
Assuming unit cell with six formula units \cite{Mai2018}, one expects $3\cdot126$ phonon branches in Ba$_5$CuIr$_3$O$_{12}$. At low temperatures 3 acoustic branches should dominate and we use the Debye model for their contribution
\begin{equation}
C_{Deb}(\theta_D,T) = 9 k_B \left(\frac{T}{\theta_D}\right) \int_0^{\theta_D/T}dx \frac{x^4 e^x}{(e^x-1)^2}.
\end{equation}
For higher temperatures the contribution of optical phonons should be included; we use the Einstein model
\begin{equation}
C_{Ein}(\omega,T) = 3 k_B \left(\frac{\hbar \omega}{T}\right)^2  \frac{e^{\frac{\hbar \omega}{T}}}{(e^{\frac{\hbar \omega}{T}}-1)^2}.
\end{equation}
We've found that a successful fitting can be performed using Einstein model with three different frequencies. In total, we have used (per formula unit)
\begin{equation}
C_{phon}(T)=C_{Deb}(\theta_D,T)/6 + 125/6( a C_{Ein}^1(\omega_1,T)+b C_{Ein}^2(\omega_2,T)+c C_{Ein}^3(\omega_3,T)),
\label{sup:eq:cphon}
\end{equation}
where $a+b+c=1$ and $\theta_D,\omega_{1,2,3}$ are fitting parameters. Their values resulting from the fit are $a=0.36\pm0.03,\;b=0.13\pm0.04,\;c=0.51\pm0.01,\;\omega_1=196\pm12\;K,\;\omega_2=97\pm9\;K,\;\omega_3=567\pm13\;K$ and the ones for the random singlet part are given in Table \ref{table:Fit}. The confidence intervals for $a,b,c$ have been calculated without taking the constraint $a+b+c=1$ into account.

\begin{table}
	\caption{\label{table:Fit}
		Fitting parameters for the phonon model \eqref{sup:eq:cphon} with respective confidence intervals. The confidence intervals for $a,b,c$ have been calculated without taking the constraint $a+b+c=1$ into account.}
	\begin{ruledtabular}
		\begin{tabular}{ccccccc}
			$a$&$b$&$c$&$\theta_D$(K)&$\hbar \omega_1$(K)&$\hbar \omega_2$ (K)&$\hbar \omega_3$ (K)\\
			\hline
			$0.36\pm0.03$&$b=0.13\pm0.04$&$0.51\pm0.01$&$74\pm15$&$196\pm12$&$97\pm9$&$567\pm13$\\
		\end{tabular}
	\end{ruledtabular}
\end{table}

\subsection{Magnetization}

\subsubsection{Obtaining $\mu_{eff}$}

\begin{figure}[h!]
	\centering
	\includegraphics[width=0.8\linewidth]{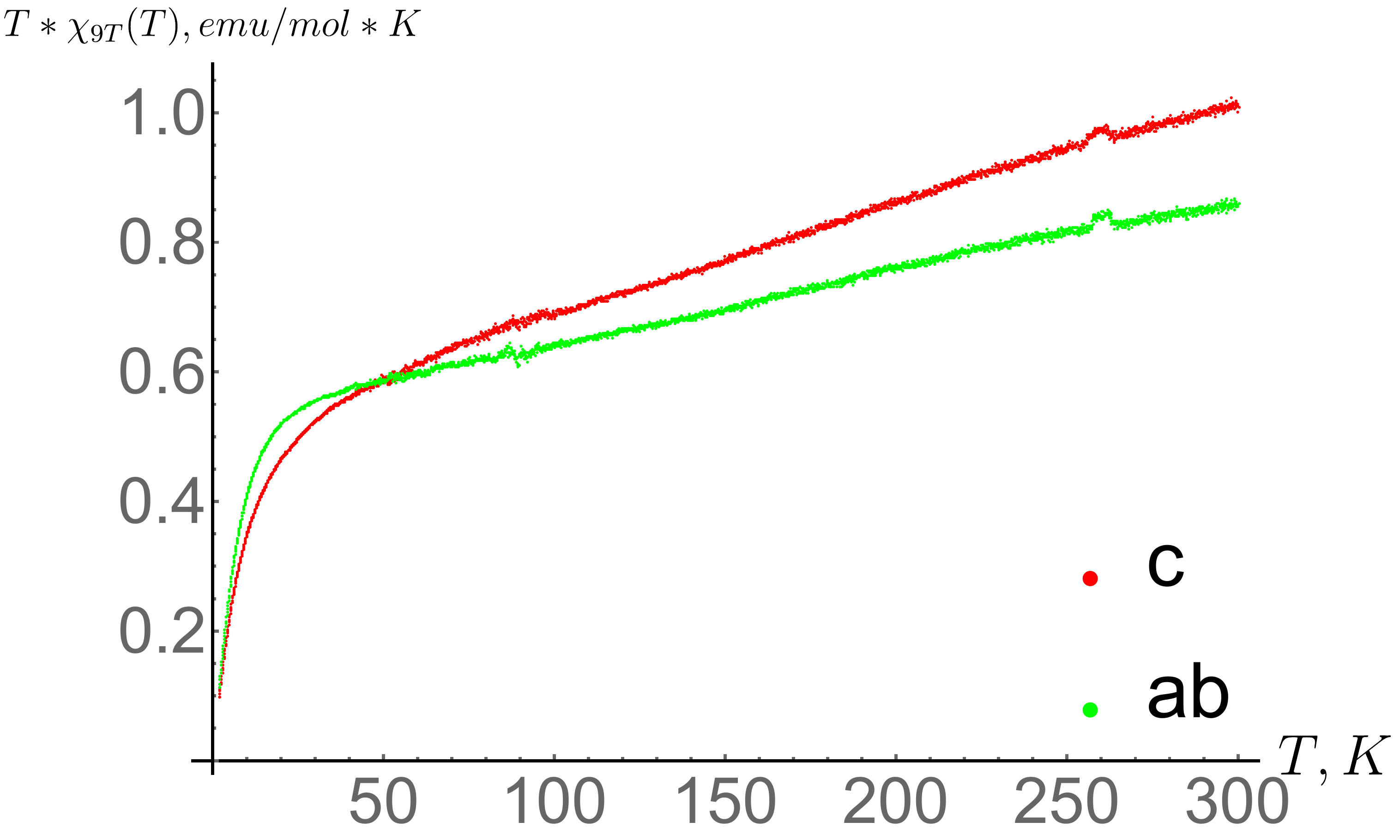}
	\caption{Susceptibility at 9T field multiplied by the temperature. The weak kink near 260\,K results from the noise of the equipment.}
	\label{sup:fig:susc}
\end{figure}

To obtain the $\mu_{eff}$ value used in the main text, we performed magnetization measurements at 9 T by using a Vibrating Sample Magnetometer (Quantum Design Dynacool)in the temperature interval 2\,K to 300\,K. The data were taken from the same sample as the one used in high-field studies. In Fig. \ref{sup:fig:susc} we present $\chi_{9T}=M(H,T)/H$.
For the susceptibility, the curves become linear in $T$ above 50 K, suggesting a Van Vleck term in addition to a Curie term. Fitting the curves between 100 and 250 K with $\chi_{9T}^{ab(c)}(T) = \frac{C^{ab(c)}}{T}+\chi_0^{ab(c)}$ we obtain: $C^{ab} = 0.518$ emu/mol*K, $\chi_0^{ab} = 0.0012$ emu/mol; $C^{c} = 0.513$ emu/mol*K, $\chi_0^{c} = 0.00174$ emu/mol. From these values one obtains $\mu_{eff}^c = 2.025 \mu_B/f.u.$ and $\mu_{eff}^{ab} = 2.035 \mu_B/f.u.$. These values are close to the effective moment expected from one Cu$^{2+}$ $\mu_\mathrm{eff}^\mathrm{Cu}=1.9\mu_B$ and one Ir trimer $\mu_\mathrm{eff}^\mathrm{Ir-tr} = 0.8\mu_B$\cite{Nguyen2019} moments per formula unit (f.u.): $\sqrt{(\mu_\mathrm{eff}^\mathrm{Cu})^2+(\mu_\mathrm{eff}^\mathrm{Ir-tr})^2} \approx2.06\mu_B$.

Note that $\chi-\chi_0$ at the lowest available temperature restricts the possible low-temperature Curie contribution due to unpaired spins to less then 20 $\%$ of the high-temperature value. Moreover, the Curie contribution is likely to be much smaller then that, as there are no signs of saturation of $T*\chi$ in Fig. \ref{sup:fig:susc} at low temperature.

\subsubsection{Calibration of the high-field measurements}
\begin{figure}[h!]
	\centering
	\includegraphics[width=0.4\linewidth]{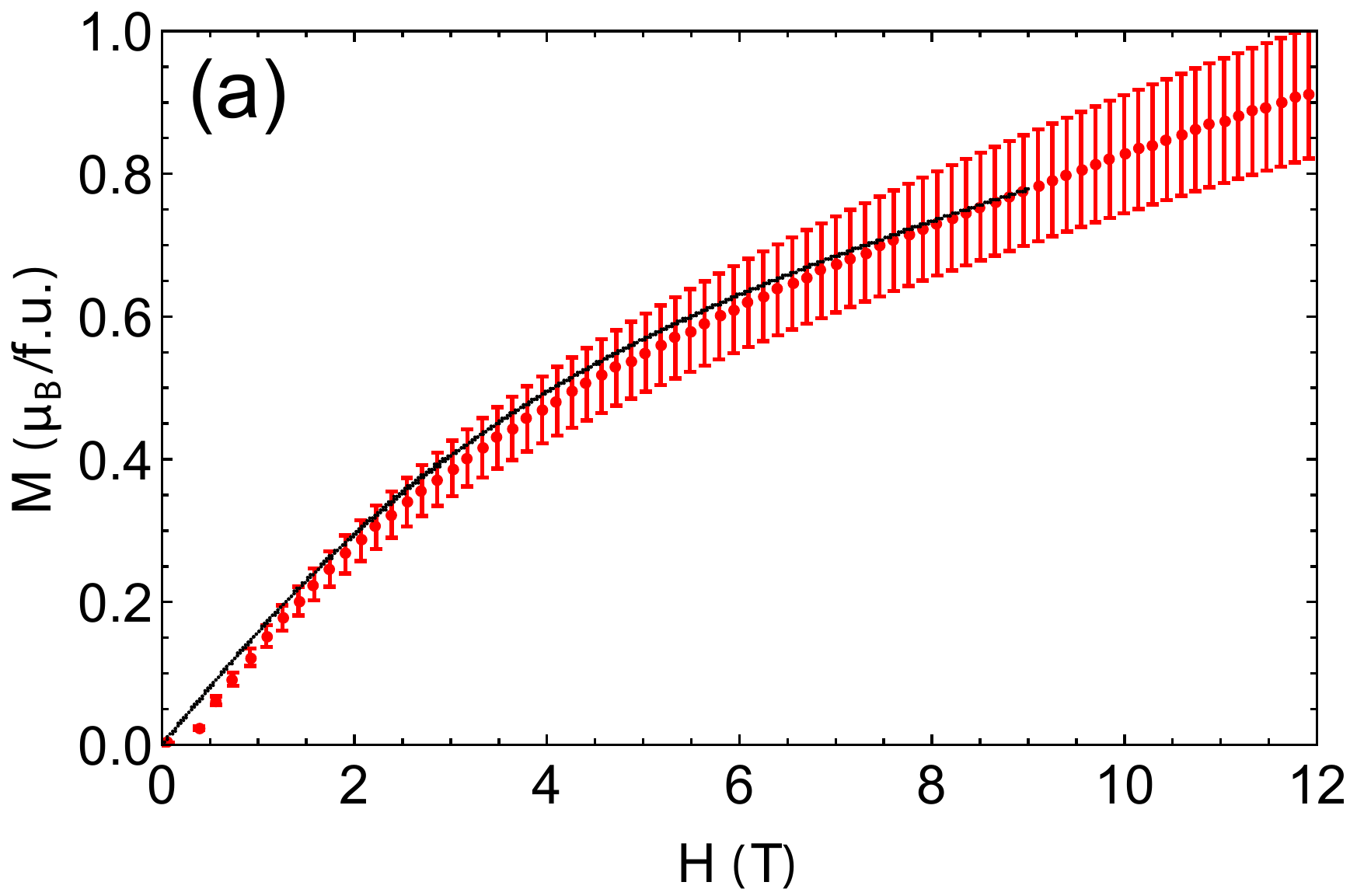}
	\includegraphics[width=0.4\linewidth]{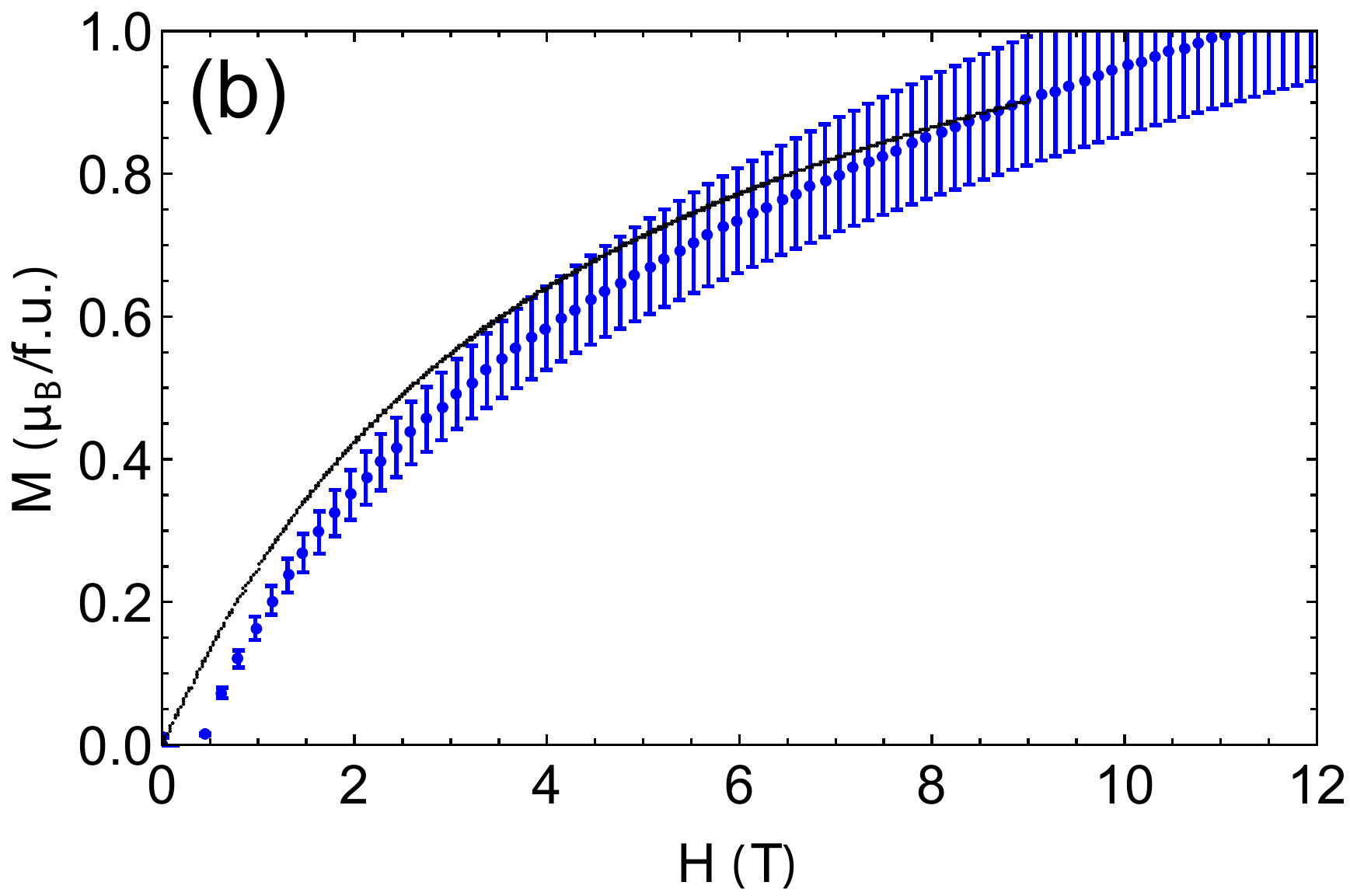}
	\caption{Pulsed-field magnetization and magnetization measured by SQUID (black) for (a) $H\parallel c$ (b) $H\parallel ab$. 10\% error bars are taken the pulsed-field data, as the sample signal has been low.}
	\label{sup:fig:magcal}
\end{figure}

The high-field measurements were calibrated with magnetization curves measured at field up to 9 T with SQUID magnetometer (Quantum Design) at 2\,K. The value of pulsed-field measurement is taken to coincide with the SQUID measurement at 9 T. The resulting curves are presented in Fig. \ref{sup:fig:magcal} with 10\% error bars for the pulsed-field data, taken for the purpose of illustration, as the sample signal has been rather low.

\begin{figure}[h!]
	\centering
	\includegraphics[width=0.8\linewidth]{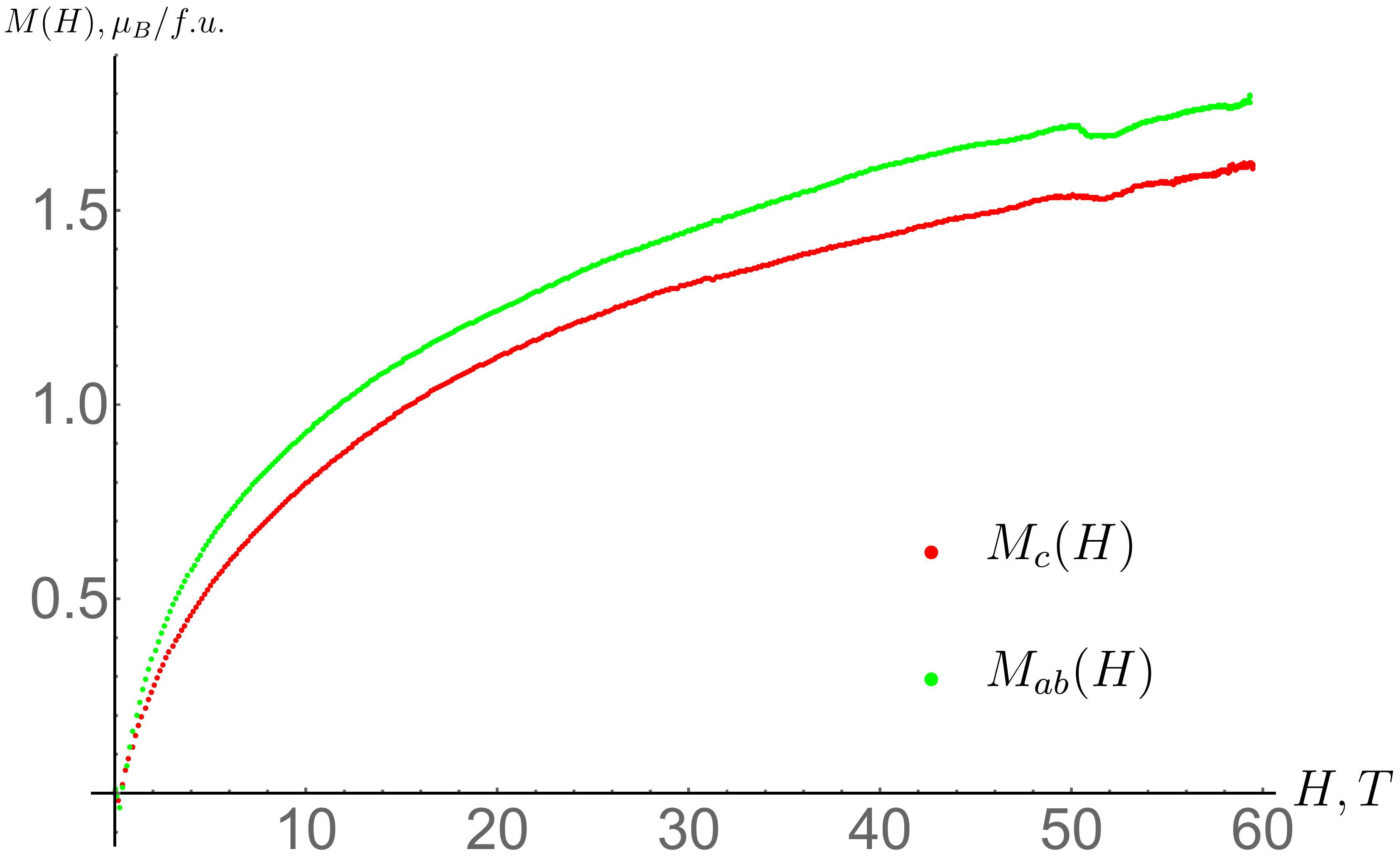}
	\caption{Magnetization curves at $T=2$ K after subtraction of the linear Van Vleck contribution.}
	\label{sup:fig:magn}
\end{figure}

One can see that the calibration curve does not fit satisfactorily the pulsed-field data for $H\parallel ab$. Furthermore, other issues appear if we proceed with magnetization along $ab$ in high-fields. In Fig. \ref{sup:fig:magn} the magnetization is presented after subtracting the Van Vleck contribution assuming it being linear in field (i.e. $M_{VV}(H) = H \chi_0$).

Because the magnetization does not saturate, the maximal value of the magnetization attained, $M_c^{max} \approx 1.62 \mu_B/f.u.$ and $M_{ab}^{max} \approx 1.79 \mu_B/f.u.$, should be smaller than the saturation moment $\mu_{sat}$. Assuming that $\mu_{eff}$ is realized two $S=1/2$ spins we have $\mu_{eff}=\sqrt{\mu_{eff,1}^2+\mu_{eff,2}^2}$, and hence $\mu_{sat} = (\mu_{eff,1}+\mu_{eff,2})\sqrt{\frac{S}{S+1}} = \frac{\mu_{eff,1}+\mu_{eff,2}}{\sqrt{3}}\leq \frac{\sqrt{2}\mu_{eff}}{\sqrt{3}}\approx 1.66 \mu_B/f.u.$ This value is smaller than $M_{ab}^{max}$. If we use the more realistic effective moments for the $Cu^{2+}$ ions ($1.9\mu_b$) and $Ir^{4+}$ trimers ($0.8\mu_b$ \cite{Nguyen2019}), $\mu_{sat} = \frac{2.7}{\sqrt{3}}\approx 1.56\mu_B/f.u.$, which is even smaller than $M_{c}^{max}$. Such discrepancy could be attributed to the issues in the pulsed-field measurements for $H\parallel ab$ evident from Fig. \ref{sup:fig:magcal}.


Given the smaller magnitude of the discrepancy for $M_{c}^{max}$, we use the $H\parallel c$ data for the analysis in the main text.


%
%
%

\end{document}